\documentclass[oribibl]{llncs}
\usepackage{url}
\usepackage{amsmath}
\usepackage{amssymb}
\usepackage{bm}
\usepackage{stmaryrd}
\usepackage{color}
\usepackage{colortbl}
\usepackage{epsf}
\usepackage{supertabular}

\usepackage{multicol}
\usepackage{smallsec}
\usepackage{times}

\newcommand{\ignore}[1]{}

\newcommand{\rred}{}
\newcommand{\re}{}
\newcommand{\bblue}{}

\newcommand{\blue}{}

\ignore{

\newtheorem{theorem}{Theorem}
\newtheorem{definition}{Definition}

\newtheorem{example}{Example}
   }

\DeclareMathAlphabet{\mathpzc}{OT1}{pzc}{m}{it}

\def\be{\begin{equation}}
\def\ee{\end{equation}}
\def\bea{\begin{eqnarray}}
\def\eea{\end{eqnarray}}
\def\nn{\nonumber}

\def\w{\wedge}
\def\ra{\rightarrow}

\def\nin{\not\in}

\def\ex{\exists}

\def\ba{\begin{align}}
\def\ea{\end{align}}
\def\bes{\begin{split}}
\def\es{\end{split}}

\def\ems{\emptyset}

\ignore{

\newcommand {\blue}[1]{\textcolor[rgb]{0.00,0.00,1.00}{#1}}

}

\newcommand{\boxtheorem}{\hfill $\Box$}
\newcommand{\nit}[1]{{\it #1}}

\newcommand{\mc}[1]{\mathcal{ #1}}

\begin{document}

\title{\vspace*{-1.5cm}{\bf On the Complexity of Query Answering under Matching Dependencies for Entity Resolution}\vspace{-3mm}}

\author{
{\bf Leopoldo Bertossi}           \hspace{2.3cm}         {\bf Jaffer Gardezi}         \\
Carleton University, SCS \hspace{0.9cm} University of Ottawa, SITE.\\
Ottawa, Canada              \hspace{2.2cm}               Ottawa, Canada\ignore{\\
jgard082@uottawa.ca \hspace{1.7cm} bertossi@scs.carleton.ca }
}

\institute{}

\maketitle
\vspace{-5mm}
\begin{abstract}
Matching Dependencies (MDs) are a relatively recent proposal for
declarative entity resolution. They are rules that
specify, given the similarities satisfied by values in
a database, what values should be considered duplicates, and have to be matched.
On the basis of a  chase-like procedure for MD enforcement, we can obtain
clean (duplicate-free) instances; actually possibly several of them. The resolved answers to queries
are those that are invariant under the resulting class of resolved instances.
In previous work we identified some  tractable cases (i.e. for certain classes of queries and MDs) of resolved query answering.
In this paper we further  investigate
the complexity of this problem, identifying some intractable cases. For a special case we obtain a dichotomy
complexity result. \vspace{-3mm}
\end{abstract}

\section{Introduction}

\vspace{-2mm}
A database may contain several representations of the same external entity. In this sense it
contains ``duplicates", which is in general considered to be undesirable. And
the database has to be cleaned. More precisely,
the problem of {\em duplicate- or entity-resolution} (ER) is about (a) detecting duplicates, and (b) merging
duplicate representations into single representations.
This is a classic and complex problem in data management, and in data cleaning in particular \cite{naumannACMCS,elmargamid,BenjellounGMSWW09}.
In this work we concentrate on the merging part of the problem, in a relational context.

A generic way to approach the problem consists in specifying what attribute values have to be
matched (made identical) under what conditions. A declarative language with a precise semantics could be used for this purpose.
In this direction, \re{matching dependencies} (MDs) have been recently introduced \cite{Fan09}.
They represent rules for
resolving pairs of duplicate representations (considering two tuples at a time). Actually, when certain similarity relationships between
attribute values hold, an MD indicates what attribute values
have to be made the same (matched).

\begin{example} \ The similarities of phone and address
indicate that the tuples refer to the same person, and the
names should be matched. Here, {\small \bblue{723-9583} $\blue{\approx}$ \bblue{(750) 723-9583}}   and   {\small \bblue{10-43 Oak St.} $\blue{\approx}$ \bblue{43 Oak St. Ap. 10}}.

\vspace{-2mm}
\begin{center}
{\small \begin{tabular}{c|c|c|c|}
$\blue{\nit{People}} \ (\blue{P})$ & Name & Phone & Address \\ \hline
&\rred{John Smith} & \bblue{723-9583} & \bblue{10-43 Oak St.} \\
&\rred{J. Smith} & \bblue{(750) 723-9583} & \bblue{43 Oak St. Ap. 10}  \\
\end{tabular}  }
\end{center}

An MD capturing this cleaning policy, could be the following:
$$\blue{P[\nit{Phone}]\approx P[\nit{Phone}]\w P[\nit{Address}]\approx P[\nit{Address}] \ \ra} \blue{P[\nit{Name}]\doteq P[\nit{Name}]}.$$
This MD involves only one database predicate, but in general,
an MD may involve two different relations. \boxtheorem
\end{example}
Here we report on new results (in Section \ref{sec:new}) on the computation of resolved query answers wrt. a set of MDs, i.e. of those answers that are invariant
under the MD-based ER process. We identify syntactic classes of MDs for which, computing resolved answers to
conjunctive queries in a syntactic class, is {\em always} intractable.

\vspace{-2mm}
\section{Preliminaries}

We assume we are dealing with relational schemas and instances.
Matching dependencies (MDs) are symbolic rules of the form: \vspace{-2mm}
\begin{equation}\label{eq:md}
\bigwedge_{i , j }R[A_i] \approx_{ij} S[B_j]  \ \ra \ \bigwedge_{k, l}R[A_k]\doteq S[B_l],\vspace{-3mm}
\end{equation}
where $R, S$ are relational predicates, and the $A_i, ...$ are attributes for them. The LHS captures similarity conditions on a pair of tuples belonging to the extensions of $\blue{R}$ and $\blue{S}$
in an instance $D$. We  abbreviate this formula as: \ $
R[{\bar A}] \approx S[{\bar B}] \ \ra \ R[\bar C]\doteq S[\bar E]
$.
MDs have a {\em dynamic interpretation} requiring that those values on the RHS
should be updated to some (unspecified) common value. Those attributes on a RHS of an MD are called \bblue{\em changeable attributes}.

The similarity predicates \bblue{$\approx$} (there may be more than one in an MD depending on the attributes involved) are treated here as built-ins, but are assumed to   satisfy:
(a) {\em symmetry}: if \bblue{$x\approx y$}, then \bblue{$y\approx x$}; and
(b) {\em equality subsumption}: if \bblue{$x = y$}, then \bblue{$x\approx y$}. However, {\em transitivity} is {\em not} assumed  (and in some application it may not hold).

\ignore{A common non-transitive similarity operator is defined in terms of
an upper bound on \bblue{edit distance} (or a related metric such as
\bblue{Affine gap distance} or \bblue{Smith-Waterman distance}) }

MDs are to be ``applied" iteratively until duplicates are solved. In order to
keep track of the changes and comparing tuples and instances, we use global
tuple identifiers, a non-changeable surrogate key for each database predicate that
has changeable attributes. The auxiliary, extra attribute (when shown) appears as the first
attribute in a relation, e.g. \bblue{$t$} is the identifier in \bblue{$R(t,\bar x)$}.
A \bblue{\em position} is a  pair \bblue{$(t,A)$}
with \bblue{$t$} a tuple id, and $A$ an attribute (of the relation where $t$ is an id).
The \blue{\em position's value}, \bblue{$t[A]$}, is  the value for $A$ in tuple
(with id) \bblue{$t$}.

A semantics for MDs acting on database instances was
proposed in \cite{front12}.
It is based on a {\em chase procedure} that is iteratively applied to the original instance \bblue{$D$}.
A \bblue{\em resolved instance}
\bblue{$D'$} is obtained from a finitely terminating
sequence
of instances, say \vspace{-2mm}
\begin{equation}
D \mapsto D_1 \mapsto D_2 \mapsto \cdots \mapsto D', \vspace{-2mm} \label{eq:seq}
\end{equation}
 terminating in $D'$,  that satisfies the MDs as  {\em equality generating dependencies} \cite{Abiteboul}, i.e.
 replacing $\doteq$ by equality.

The semantics specifies the one-step transitions  or updates allowed to go from \bblue{$D_{i-1}$} to \bblue{$D_i$}, i.e. ``$\blue{\mapsto}$" in (\ref{eq:seq}).
Only \bblue{\em modifiable positions} within the instance are allowed to change their values in such a step,
and as forced by the MDs. Actually, the
modifiable positions syntactically depend on a whole set $M$ of MDs and instance at hand; and can be recursively defined (see \cite{front12,datalog12}
for the details). Intuitively, a position $\blue{(t,A)}$ is modifiable iff: (a)
There is a $\blue{t'}$ such that $\blue{t}$ and $\blue{t'}$ satisfy the
similarity condition of an MD with $A$ on the RHS; or (b) $\blue{t[A]}$ has not already been resolved (it is
different from one of its other duplicates).

\begin{example} \ Consider the MD \ \bblue{$R[A] = R[A] \ra R[B]\doteq R[B]$}, and the instance $R(D)$ below.
The positions of the underlined values in $\blue{D}$ are modifiable, because
their values are unresolved (wrt the MD).

\begin{multicols}{2}
\begin{center}
\blue{\begin{tabular}{l|c|c|}
$R(D)$ & $A$ & $B$ \\ \hline
$t_1$ & $a$ & \underline{$b$} \\
$t_2$ & $a$ & \underline{$c$}
\end{tabular}}
$~~~\blue{\mapsto}~~~$
\blue{\begin{tabular}{l|c|c|}
$R(D')$ & $A$ & $B$ \\ \hline
$t_1$ & $a$ & $\blue{d}$ \\
$t_2$ & $a$ & $\blue{d}$
\end{tabular}}
\end{center}

\noindent $\blue{D'}$ is a \bblue{resolved instance} since it satisfies the
MD interpreted as an FD (the update value \bblue{$d$} is arbitrary).
\end{multicols}

\noindent $\blue{D'}$ has no modifiable positions with unresolved values:
the values for $B$ are already the
same, so there is no reason to change them. \boxtheorem
\end{example}
More formally, the {\em single step semantics} is a follows.
Each pair $\blue{D_i, D_{i+1}}$ in an update
sequence (\ref{eq:seq}), i.e. a chase step, must {\em satisfy} the set $M$ of MDs, modulo unmodifiability,
denoted \ $(D_i,D_{i+1}) \models_{\it um} M$, which   holds iff: \ (a)
 For every MD, say \ $\blue{R[\bar A]\approx S[\bar B]\ra R[\bar C]\doteq S[\bar D]}$
\ and pair of tuples $\blue{t_R}$ and $\blue{t_S}$,
if $\blue{t_R[\bar A]\approx t_S[\bar B]}$ in $\blue{D_i}$, then $\blue{t_R[\bar C] = t_S[\bar D]}$
in $\blue{D_{i+1}}$; and (b)
 The value of a position can only differ between
$\blue{D_i}$ and $\blue{D_{i+1}}$ if it is modifiable wrt $\blue{D_i}$.

This semantics stays as close as possible to the spirit of
the MDs as originally introduced \cite{Fan09}, and also {\em uncommitted} in the sense that the MDs do not specify
how the matchings have to be realized.\footnote{We have proposed and investigated other semantics. One of them is as
above, but with a modified chase conditions, e.g. applying one MD at a time. Another one imposes that previous resolutions cannot be unresolved.
In \cite{icdt11,tocs,kr12} a semantics that uses {\em matching functions} to choose a value for a match is developed.}

\begin{example} \label{ex:two} \ Consider the following instance and set of MDs. Here, attribute $\blue{R(C)}$ is changeable. Position \blue{$(t_2,C)$ is not modifiable} wrt. $\blue{M}$
and $\blue{D}$: There is no justification
\begin{multicols}{2}
\begin{center}
\hspace*{1cm}\begin{tabular}{l|c|c|c|}
$R(D)$ & $A$ & $B$ & $C$ \\ \hline
$t_1$ &$a$ & $b$ & $d$ \\
$t_2$ &$a$ & $c$ & $\underline{e}$ \\
$t_3$ &$a$ & $b$ & $e$
\end{tabular}
\end{center}

\phantom{mmmmmmmm}

\vspace*{-12mm}
\begin{eqnarray}
R[A] = R[A] &\ra& R[B]\doteq R[B]\nn\\
R[B] = R[B] &\ra& R[C]\doteq R[C]. \nn
\end{eqnarray}
\noindent
\end{multicols}

\noindent
to change its value {\em in one step} on the basis of an MD and $\blue{D}$. However,
position \blue{$(t_1,C)$ is modifiable}. We obtain two resolved instances for $\blue{D}$: \ \  $\blue{D_1}$ and $\blue{D_2}$ below.

\begin{multicols}{2}
{\small
\begin{center}
\begin{tabular}{l|c|c|c|}
$R(D_1)$ & $A$ & $B$ & $C$ \\ \hline
$t_1$ &$a$ & $b$ & $d$ \\
$t_2$ &$a$ & $b$ & $d$ \\
$t_3$ &$a$ & $b$ & $d$
\end{tabular}
~~~~~~\begin{tabular}{l|c|c|c|}
$R(D_2)$ & $A$ & $B$ & $C$\\ \hline
$t_1$ & $a$ & $b$ & $e$\\
$t_2$ & $a$ & $b$ & $e$\\
$t_3$ & $a$ & $b$ & $e$
\end{tabular}
\end{center} }

\noindent $\blue{D_1}$ cannot be obtained in a single (one step) update since
the underlined value is for a non-modifiable position. However,  $\blue{D_2}$ can. \boxtheorem
\end{multicols}
\end{example}
\vspace{-4mm}Among the {\em resolved instances} we prefer those that are  closest to the original instance.
Accordingly, a {\em minimally resolved instance} (MRI) of $\blue{D}$ is a resolved
instance $\blue{D'}$ such that {\em the number of changes of attribute values} comparing $\blue{D}$ with $\blue{D'}$ is
a minimum. In Example \ref{ex:two}, instance $\blue{D_2}$ is an MRI, but not $\blue{D_1}$ \ (2 vs. 3 changes).
We denote with $\nit{Res}(D,M)$ and $\nit{MinRes}(D,M)$ the classes of resolved, resp. minimally resolved, instances of $D$
wrt $M$.

Given a conjunctive query \bblue{$\mc{Q}$}, a set of MDs \bblue{$M$}, and an
instance \bblue{$D$}, the {\em resolved answers} to $\mc{Q}$ from $D$ are those that are invariant under the
entity resolution process, i.e.
they are answers to \bblue{$\mc{Q}$} that are true in all
MRIs of $\blue{D}$: \ $\nit{ResAns}_M(\mc{Q},D) := \{\bar{c}~|~D' \models \mc{Q}[\bar{c}], \mbox{ for every } D' \in \nit{MinRes}(D,M)\}$.
We denote with $\nit{RA}(\mathcal{Q},M)$ the decision problem $\{(D,\bar{c})~|~ \bar{c} \in \nit{ResAns}_M(\mc{Q},D)\}$.

The definition of resolved answer is reminiscent of that of consistent query answers (CQA) in databases that may not
satisfy given integrity constraints (ICs) \cite{Arenas99,B2006}. Much research in CQA has been about developing (polynomial-time) query rewriting methodologies.
The idea is to rewrite a query, say conjunctive, into a new query such that the new query on the inconsistent database returns as usual answers the consistent
answers to the original query. In all the cases identified in the literature on CQA (see \cite{bertossi11} for a survey, and \cite{wijsenTods12} for recent results) depending on the class of conjunctive
query and ICs involved, the rewritings that produce polynomial time CQA have been first-order.
Doing something similar for resolved query answering (RQA) under MDs brings new challenges: (a)  MDs contain the non-transitive
similarity predicates. (b) Enforcing consistency of updates requires computing the transitive
closure of such operators. (c) The minimality of {\em value changes} (that is not always used in CQA or considered for consistent rewritings). (d) The semantics of resolved query answering for MD-based
entity resolution
is given, in the end, in terms of a chase procedure.\footnote{For some implicit connections between repairs and chase procedures, e.g. as used in data exchange see
\cite{kolaitisICDT12}, and as used under database completion with ICs see \cite{cali03}.} However, the semantics of CQA is model-theoretic, given in terms repairs that are not operationally defined, but arise
from set-theoretic conditions.\footnote{For additional discussions of differences and connections between CQA and resolved query answering see  \cite{front12,sum12}.}

\vspace{-3mm}
\section{Tractability and Datalog Query Rewriting}

\vspace{-1mm}
In \cite{datalog12,sum12}, a query rewriting methodology for RQA under MDs was presented. In this case,
the rewritten queries turn out to be Datalog
queries with counting, and can be obtained for two main classes of sets of MDs: (a)   MDs do not depend on each other, i.e. {\em non-interacting} sets of MDs \cite{front12};
(b) MDs depend cyclically on each other, e.g. a set containing
$\blue{R[A]\approx R[A]\ra R[B]\doteq R[B]}$ and
$\blue{R[B]\approx R[B]\ra R[A]\doteq R[A]}$ (or relationships like this by transitivity).

Here cycles help us, because the termination
condition for the chase imposes a simple form on the minimally resolved instances
(easier to capture and characterize) \cite{datalog12}.
For these sets of MDs a conjunctive query can be rewritten
to retrieve, in polynomial time, the resolved answers, provided
there are no joins on existentially quantified
variables corresponding to changeable attributes: {\em unchangeable
attribute join conjunctive}  (\bblue{UJCQ}) queries \cite{sum12}.
For example, for the MD $R[A] = R[A]\ra R[B,C]\doteq R[B,C]$ on schema $R[A,B,C]$,
$\mathcal{Q}\!:
\ex x \ex y \ex z(R(x,y,c)\w R(z,y,d))$ is {\em not} UJCQ; whereas
$\mathcal{Q}'\!:
\ex x \ex z(R(x,y,z)\w R(x,y',z')$ is UJCQ. For queries outside UJCQ, the resolved answer problem can be intractable even for
one MD \cite{sum12}.

The case of a set of MDs consisting of \vspace{-2mm}
\begin{equation}
R[A]\approx R[A]\ra R[B]\doteq R[B] \mbox{ and } R[B]\approx R[B]\ra R[C]\doteq R[C], \label{eq:lid} \vspace{-2mm}
\end{equation}
which is neither non-interacting nor cyclic, is
not covered by the positive cases for Datalog rewriting above. Actually, for this set RQA becomes intractable for very
simple queries, like $\mathcal{Q}(x, z)\!: \exists y R(x, y, z)$, that is UJCQ \cite{front12}.

\vspace{-3mm}
\section{Intractability of Computing Resolved Query Answers}\label{sec:new}

\vspace{-1mm}
In the previous section we briefly described classes of queries and MDs for which RQA can be done in
polynomial time in data (via the Datalog rewriting). We also showed that there are intractable cases, by pointing to a specific query and
set of MDs. The questions that naturally arise are: (a) What happens outside the Datalog rewritable cases in terms of complexity of
RQA? \ (b) The exhibited query and MDs correspond to a more general pattern for which intractability holds? We address these questions here.

For all sets $M$ of MDs we consider below, at most two relational predicates appear in $M$, and when there are two
predicates, both appear in all MDs  in $M$. According to the syntactic restrictions for MDs in (\ref{eq:md}),
 those two predicates occur in all conjuncts of an MD in $M$. Furthermore, all the sets of MDs considered below will turn out to
 be, as previously announced, both interacting and acyclic. Both notions and others can be captured in terms of the MD {\em graph},
 $\nit{MDG}(M)$, a directed graph, such that, for $m_1, m_2 \in M$, there is an edge from $m_1$ to $m_2$ if there is
an overlap between $\nit{RHS}(m_1)$ and $\nit{LHS}(m_2)$ (the right- and left-hand sides of the arrows as sets of attributes) \cite{front12}.
 $M$ is acyclic when $\nit{MDG}(M)$ is acyclic.
Our results require several terms and notation that we now define.

\vspace{-1mm}
\begin{definition}\label{def:rquery} \em 
Consider a set $M$ of MDs involving the predicates $R$ and $S$.
A {\em changeable attribute query} $\mc{Q}$ is a (conjunctive) query  in UJCQ, containing a conjunct
of the form $R(\bar x)$ or $S(\bar y)$ with  all variables free. Such
a conjunct is called a {\em free occurrence} of
the predicate $R$ or $S$.\boxtheorem
\end{definition}
By definition, the class of {\em changeable attribute queries} (CHAQ) is a subclass of UJCQ. Both classes depend on
the set of MDs at hand. For example, for the MDs in (\ref{eq:lid}), $\exists y R(x, y, z) \in \mbox{ UJCQ } \smallsetminus \mbox{ CHAQ}$, but  
$\exists w \exists t(R(x, y, z) \wedge S(x,w, t)) \in \mbox{ CHAQ}$.
We confine attention to UJCQ and subsets of it because, as mentioned in the previous
section, intractability limits the applicability of the duplicate resolution method for queries outside UJCQ.
The requirement that the query contains a free occurrence
of $R$ or $S$ eliminates from consideration certain queries in UJCQ for which the resolved answer
problem is trivially tractable. For example, for MDs in (\ref{eq:lid}), the query $\ex y \ex z R(x,y,z)$ is not in CHAQ, but is tractable simply because it does not return the values of a changeable attribute (the resolved answers are the answers in the usual sense).

\vspace{-1mm}
\begin{definition}\label{def:hard} \em 
A set $M$ of MDs is {\em hard} if for every
CHAQ $\mathcal{Q}$,  $\nit{RA}(\mathcal{Q},M)$
is \nit{NP}-hard. $M$ is {\em easy} if for every CHAQ $\mathcal{Q}$,  $\nit{RA}(\mathcal{Q},M)$ is in \nit{PTIME}.
\boxtheorem
\end{definition}
Of course, a  set of MDs may not be hard or easy. In the following we give some syntactic conditions that guarantee
hardness for classes of MDs. 
\begin{definition} \em 
Let $m$ be an MD. The symmetric binary
relation $\nit{LRel}(m)$ ($\nit{RRel}(m)$) relates each pair of attributes
$A$ and $B$ such that a conjunct of the form $R[A]\approx S[B]$ (resp. $R[A]\doteq S[B]$)
appears in $\nit{LHS}(m)$ (resp. $\nit{RHS}(m)$). An {\em L-component}
({\em R-component}) of $m$ is an equivalence
class of the reflexive and transitive closure, $\nit{LRel}(m)^\nit{eq}$ (resp. $\nit{RRel}(m)^\nit{eq}$), of
$\nit{LRel}(m)$ (resp. $\nit{RRel}(m)$). \boxtheorem
\end{definition}

The first results concern {\em linear pairs} of MDs, i.e.  those whose graph $\nit{MDG}(M)$
consisting of the vertices $m_1$ and $m_2$, say \vspace{-2mm}
\bea
\hspace*{-4mm}m_1\!\!: R[\bar A]\approx_1 S[\bar B]\ra R[\bar C]\doteq S[\bar E], \mbox{ and }
m_2\!\!: R[\bar F]\approx_2 S[\bar G]\ra R[\bar H]\doteq S[\bar I],   \label{eq:mds} \vspace{-5mm}
\eea
with only
an edge from $m_1$ to $m_2$, i.e. $(R[\bar C] \cup S[\bar E]) \cap (R[\bar F] \cup S[\bar G]) \neq \emptyset$, whereas
$(R[\bar H] \cup S[\bar I])  \cap (R[\bar A] \cup S[\bar B]) = \emptyset$.
The linear pair is denoted by $(m_1,m_2)$.

\begin{definition}\label{def:equivset} \em 
Let $(m_1,m_2)$ be a linear pair as in (\ref{eq:mds}).
\ignore{\bea
m_1\!:~R[\bar A]\approx_1 S[\bar C]\ra R[\bar E]\doteq S[\bar F],\nn\\
m_2\!:~R[\bar G]\approx_2 S[\bar H]\ra R[\bar I]\doteq S[\bar J].  \nn
\eea}
(a)  $B_R$ is a binary (reflexive and symmetric) relation on
attributes of $R$: \ $(R[U_1],R[U_2]) \in
B_R$ iff $R[U_1]$ and $R[U_2]$ are in the same
R-component of $m_1$ or the same L-component of $m_2$.
Similarly for $B_S$.

\noindent (b) An {\em $R$-equivalent set} ($R$-ES) of attributes of $(m_1,m_2)$
is an equivalence class of $\nit{TC}(B_R)$, the transitive closure of $B_R$, with at least one attribute in the equivalence class
belonging to $\nit{LHS}(m_2)$. The definition of an {\em $S$-equivalent set} ($S$-ES) is the same, with $R$ replaced by $S$.

\noindent (c) An ($R$ or $S$)-ES $E$ of $(m_1,m_2)$ is {\em bound} if
$E\cap  \nit{LHS}(m_1)$ is non-empty. \boxtheorem
\end{definition}

\begin{theorem}\label{thm:combine} \em 
Let $(m_1,m_2)$ be a linear pair as in (\ref{eq:mds}), with $R$ and $S$ distinct predicates.
Assume that each similarity relation has an infinite
set of mutually dissimilar elements. \
Let $E_R$ and
$E_S$ be
the classes of $R$-ESs and  $S$-ESs, resp.  \
The pair $(m_1,m_2)$ is hard
if $\nit{RHS}(m_1) \cap \nit{RHS}(m_2) = \ems$, and at least one of the
following {\em does not} hold: \vspace{-2mm}
\begin{itemize}
\item[(a)] At least one of the following is true: (i) there are no
attributes of $R$ in $\nit{RHS}(m_1)\cap\nit{LHS(m_2)}$; (ii) all ESs in $E_R$ are bound; or (iii)
for each L-component $L$ of $m_1$, there is an attribute
of $R$  in $L \cap \nit{LHS}(m_2)$.
\item[(b)] At least one of the following is true: (i) there are no
attributes of $S$ in $\nit{RHS}(m_1)\cap\nit{LHS(m_2)}$; (ii) all ESs in $E_S$ are bound; or (iii)
for each L-component $L$ of $m_1$, there is an attribute
of $S$ in $L \cap \nit{LHS}(m_2)$.
\boxtheorem
\end{itemize}
\end{theorem}
\vspace{-2mm}Theorem \ref{thm:combine} says that a linear pair of MDs is hard unless
the syntactic form of the MDs is such that
there is a certain association between changeable attributes in $\nit{LHS}(m_2)$ and attributes
in $\nit{LHS}(m_1)$ as specified by conditions (ii) and (iii). When $m_1$ is
applied to an instance, similarities can be produced among the values of
attributes of $\nit{RHS}(m_1)$ which are not required by the chase but
result from a particular choice of update values. Such {\em accidental similarities}
affect the subsequent updates made by applying $m_2$, making the query
answering problem intractable \cite{front12}. For pairs of
MDs satisfying (a)(ii) or (a)(iii) (or (b)(ii) or (b)(iii)) in Theorem \ref{thm:combine}, the
similarities resulting from applying $m_2$ are restricted to a subset of
those that are already present among the values of attributes
in $\nit{LHS}(m_1)$, making the problem tractable.

However, when condition
(ii) or (iii) is satisfied, accidental similarities among the values
of attributes in $\nit{RHS}(m_1)$ cannot be passed on to values of
attributes in $\nit{RHS}(m_2)$.

This result gives a syntactic condition for hardness.
It is an important result, because it applies to many cases of
practical interest.
For example, the linear pair  $(m_1,m_2)$ in (\ref{eq:lid})
turns out to be hard (for all CHAQ queries, in addition to $\exists y R(x, y, z)$).

All syntactic conditions/constructs on attributes above, in particular, the transitive closures on attributes, are
``orthogonal" to semantic properties of the similarity relations. When similarity predicates are transitive, every linear pair not
satisfying the hardness criteria of Theorem \ref{thm:combine} is easy.

\begin{theorem} (dichotomy for transitive similarity) \em Let $(m_1, m_2)$ be a linear pair with $\nit{RHS}(m_1) \cap \nit{RHS}(m_2) = \emptyset$.  If the
similarity
operators are transitive, then $(m_1, m_2)$ is either
easy or hard. \boxtheorem
\end{theorem}
The next result concerns {\em pair-preserving} acyclic sets of  MDs, defined by: \  $M$ is pair-preserving if,
for any attribute $R[A]$ occurring in a MD, there is
only one attribute $S[B]$ such that $R[A]\approx S[B]$
or $R[A]\doteq S[B]$ occur in an MD. These sets of MDs can be of arbitrary size (
still subject to the condition of containing at most two predicates). The pair-preserving assumption typically holds in a duplicate resolution
setting, since the values of pairs of attributes are normally
compared only if they hold the same type of information
(e.g. they are both addresses or both names).

\begin{definition}\label{def:inclusion} \em 
Let $M$ be  pair-preserving and acyclic,
$B$ an attribute in $M$, and
$M' \subseteq M$. \  $B$ is {\em non-inclusive}
wrt. $M'$ if, for every $m \in M \! \smallsetminus \! M'$
with $B\in \nit{RHS}(m)$, there is an
attribute $C$ such that: \
(a) $C\in \nit{LHS}(m)$, \
(b) $C\nin \bigcup_{m'\in M'}\nit{LHS}(m')$, and (c) $C$ is {\em non-inclusive}
wrt. $M'$. \  \ignore{An attribute that is not non-inclusive wrt. $M'$ is
{\em inclusive} wrt. $M'$.}
\boxtheorem
\end{definition}
This is a recursive definition of non-inclusiveness. The base case occurs when $C$ is not in $\nit{RHS}(m)$
for any $m$, and so must be inclusive (i.e. not non-inclusive). Because $C\in \nit{LHS}(m)$ in the definition, for any $m_1$ such that $C\in \nit{RHS}(m_1)$, there is an edge from $m_1$ to $m$. Therefore, we are traversing an edge backwards with each recursive step, and the recursion terminates by the acyclicity assumption.

Non-inclusiveness is a generalization of conditions (a) (iii) and (b) (iii)
in Theorem \ref{thm:combine} to a set of arbitrarily many MDs. It expresses
a condition of inclusion of attributes in the left-hand side of one MD
in the left-hand side of another. Theorem \ref{thm:main} tells us that
a set of MDs that is non-inclusive in this sense is hard. Notice that
the condition of Theorem \ref{thm:combine} that there exists an ES that is not bound
does not appear in Theorem \ref{thm:main}. This is because, by
the pair-preserving requirement, there cannot be a bound ES for
any pair of MDs in the set. For linear pairs, Theorem \ref{thm:main} becomes Theorem \ref{thm:combine}.

\begin{theorem}\label{thm:main} \em 
Let $M$ be pair-preserving and acyclic. Assume
there is $\{m_1,m_2\}$ $\subseteq  M$, and attributes $C\in\nit{RHS}(m_2)$,
$B\in\nit{RHS}(m_1)\bigcap\nit{LHS}(m_2)$ with: \ (a) $C$ is non-inclusive
wrt $\{m_1,m_2\}$, and (b) \ $B$ is non-inclusive wrt $\{m_2\}$.
Then, $M$ is hard.
\boxtheorem
\end{theorem}
\vspace{-1mm}\noindent {\small {\bf Acknowledgments:} \ Research supported by the NSERC Strategic
Network on Business Intelligence (BIN ADC05), NSERC/IBM CRDPJ/371084-2008, and NSERC Discovery.}

\end{document}